\documentclass[twocolumn]{aastex631}

\usepackage{amsmath}
\usepackage{amssymb}
\usepackage{braket}
\usepackage{amssymb}
\usepackage{color}
\usepackage{xcolor}
\usepackage{graphicx}
\usepackage{latexsym}
\usepackage{listings}
\usepackage{mathrsfs}
\usepackage{letltxmacro}
\usepackage[normalem]{ulem}

\usepackage{subfigure}

\usepackage{verbatim}
\usepackage{tabularx}
\usepackage{ragged2e}
\usepackage{multirow}
\usepackage{amsbsy}

\usepackage{numprint}
\usepackage{makecell}
\usepackage[utf8]{inputenc}

\newcommand{\R}{\mathbb{R}}
\newcommand{\N}{\mathbb{N}} 

\newcommand{\V}{\mathcal{V}}

\newcommand{\F}{\mathcal{F}}

\newcommand{\M}{\mathcal{M}}

\newcommand{\I}{\mathcal{I}}

\newcommand{\B}{\text{B}}

\npdecimalsign{.}
\npthousandsep{,}

\newcolumntype{Y}{>{\RaggedRight\arraybackslash}X}

\graphicspath{{./figures/}}

\sloppypar

\definecolor{darkperiwinkle}{RGB}{153 153, 192}

\received{}
\revised{}
\accepted{}
\submitjournal{ApJL}

\graphicspath{{./}{figures/}}

\begin{document}

\title{Outflows from Short-Lived Neutron-Star Merger Remnants Can Produce a Blue Kilonova}
\shorttitle{Blue Kilonovae from NS Remnants}
\shortauthors{Curtis et al.}

\correspondingauthor{Sanjana Curtis}
\email{sanjanacurtis@uchicago.edu}

\correspondingauthor{Pablo Bosch}
\email{p.boschgomez@uva.nl}

\author[0000-0002-3211-303X]{Sanjana Curtis}
\affiliation{Department of Astronomy and Astrophysics, University of Chicago, Chicago  IL 60637}
\affiliation{Anton Pannekoek Institute for Astronomy, Institute of
High-Energy Physics, and Institute of Theoretical Physics, University of Amsterdam,
Science Park 904, 1098 XH Amsterdam, The Netherlands}

\author[0000-0002-3211-303X]{Pablo Bosch}
\affiliation{Anton Pannekoek Institute for Astronomy, Institute of
High-Energy Physics, and Institute of Theoretical Physics, University of Amsterdam,
Science Park 904, 1098 XH Amsterdam, The Netherlands}

\author[0000-0002-3211-303X]{Philipp M{\"o}sta}
\affiliation{Anton Pannekoek Institute for Astronomy, Institute of
High-Energy Physics, and Institute of Theoretical Physics, University of Amsterdam,
Science Park 904, 1098 XH Amsterdam, The Netherlands}

\author[0000-0001-6982-1008]{David Radice}
\affiliation{Institute for Gravitation and the Cosmos, The Pennsylvania State University, University Park, PA 16802, USA}
\affiliation{Department of Physics, The Pennsylvania State University, University Park, PA 16802, USA}
\affiliation{Department of Astronomy \& Astrophysics, The Pennsylvania State University, University Park, PA 16802, USA}

\author[0000-0002-2334-0935]{Sebastiano Bernuzzi}
\affiliation{Theoretisch-Physikalisches Institut,
  Friedrich-Schiller-Universit{\"a}t Jena, 07743, Jena, Germany}
  
\author[0000-0002-0936-8237]{Albino Perego}
\affiliation{Dipartimento di Fisica, Università di Trento, Via Sommarive 14, 38123 Trento, Italy}
\affiliation{NFN-TIFPA, Trento Institute for Fundamental Physics and Applications, ViaSommarive 14, I-38123 Trento, Italy}

\author[0000-0002-4518-9017]{Roland Haas}
\affiliation{National Center for Supercomputing Applications, University of Illinois, 1205 W Clark St, Urbana, Illinois, USA}
\affiliation{Department of Physics, University of Illinois, 1110 West Green St, Urbana, Illinois, USA}

\author[0000-0002-4518-9017]{Erik Schnetter}
\affiliation{Perimeter Institute for Theoretical Physics, Waterloo, Ontario, Canada}
\affiliation{Department of Physics and Astronomy, University of Waterloo, Waterloo, Ontario, Canada}
\affiliation{Center for Computation \& Technology, Louisiana State University, Baton Rouge, Louisiana, USA}

\begin{abstract}
We present a 3D general-relativistic magnetohydrodynamic simulation of a short-lived neutron star remnant formed in the aftermath of a binary neutron star merger. The simulation uses an M1 neutrino transport scheme to track neutrino-matter interactions and is well-suited to studying the resulting nucleosynthesis and kilonova emission. We find that the ejecta in our simulations under-produce $r$-process abundances beyond the second $r$-process peak. For sufficiently long-lived remnants, these outflows \textit{alone} can produce blue kilonovae, including the blue kilonova component observed for AT2017gfo.

\end{abstract}


\section{Introduction} 

The observation of AT2017gfo \citep{Coulter2017, SoaresSantos2017, Arcavi2017} -- a kilonova arising from merging neutron stars -- has provided strong evidence that such mergers are a site where heavy elements are produced via the rapid neutron-capture process or $r$-process \citep{Pian2017, Kasen2017Nature}. The optical and infrared spectra of this transient show an early `blue' peak at ultraviolet/optical frequencies and a late `red' peak in the near-infrared  \citep{Evans2017, McCully2017, Nicholl2017, Tanvir2017, Chornock2017}. This behavior is thought to arise from the presence of two or more distinct outflow components, which differ with respect to their total mass as well as velocities and opacities \citep{Perego2017, Cowperthwaite2017, Drout2017, Villar2017}. 

The neutron-richness of the ejecta is an important quantity for nucleosynthesis, indicated by its electron fraction $Y_e = N_P/N_B$, where $N_P$ is the total number of protons and $N_B$ is the total number of baryons. In general, relatively neutron-rich ejecta ($Y_e \lesssim 0.25$) synthesize a significant mass-fraction of heavy elements including the lanthanides (140 $\leq A < 176$). Lanthanides produce strong optical line blanketing and shift emission towards infrared bands, thus giving rise to a `red' kilonova. Less neutron-rich ejecta ($Y_e \gtrsim 0.25$) are lanthanide-poor with a correspondingly lower opacity and produce a `blue' kilonova. 

While the multi-component fits to AT2017gfo provide reasonable estimates of the ejecta properties, linking them to various outflow components found in merger simulations has been challenging \citep{Nedora2021}. The red component is thought to arise from the tidal dynamical ejecta and/or accretion disk outflows while the exact origin of the blue component remains unknown. The blue component will be the focus of this paper.

The ejecta properties derived for the blue component in \cite{Villar2017} include a total mass of $\sim 2 \times 10^{-2} M_{\odot}$, high velocities $\sim 0.27c$, and a low opacity compatible with a composition characterized by $Y_e \sim 0.25-0.35$. Shock-heated dynamical ejecta in the polar region may have the requisite velocities and composition but are insufficiently massive. Accretion disk outflows may produce massive ejecta with the appropriate composition but with much lower velocities. The production of the blue component thus requires the presence of another mechanism.

Several possibilities have been proposed to simultaneously fit these estimates, including spiral-wave winds \citep{Nedora2019}, and neutrino-heated, magnetically-accelerated outflows from a strongly magnetized neutron star (NS) merger remnant \citep{Metzger2018}. The latter suggestion motivated the modeling of a NS remnant with 3D GRMHD simulations (including neutrinos via a leakage scheme) in \cite{Moesta2020}, who found the outflow properties to be broadly in agreement with those estimated for the blue component. An investigation of the resulting emission was carried out in \cite{Curtis2023}, who found that these outflows produce a red kilonova that peaks on the timescale of a day.  

Recently, \cite{Most:23a} showed that NS merger remnant outflows can produce flares, jets, and quasi-periodic outflows, and \cite{Combi:23} argued that disk winds dominate the kilonova emission, with the NS remnant only being responsible for a blue precursor signal. 
All of these simulations treated the neutrino-matter interactions using a leakage scheme or one-moment schemes, which do not accurately capture the $Y_e$ evolution. \cite{Ciolfi2020} did not include neutrino radiation in their simulations but followed over 250 ms of post-merger evolution, concluding that the magnetically driven baryon wind can produce an ejecta component as massive and fast as required by the observed blue kilonova.

Here, we present 3D GRMHD simulations of a magnetized short-lived NS merger remnant using a two-moment M1 scheme for neutrino transport. We find an outflow that reaches quasi-steady-state operation in agreement with \cite{Moesta2020}. We calculate $r$-process abundances in the ejecta and predict kilonova light curves. The ejecta in our simulations do not produce a robust $r$-process pattern beyond the first $r$-process peak. Additionally, we find that given a sufficiently long-lived NS remnant, such outflows \textit{alone} can produce blue kilonovae, including the blue kilonova component observed for AT2017gfo. 

The paper is organized as follows. In Section \ref{sec:methods}, we describe our input models as well as numerical methods and codes used to carry out the simulations and compute abundances and light curves. We present the outflow properties in Section \ref{sec:hmns:outflows}, the ejecta composition in Section \ref{sec:hmns:r-process}, and kilonova light curves in Section \ref{sec:hmns:kn}. We discuss the implications of our results and future directions in Section \ref{sec:summary}.

\section{Methods}
\label{sec:methods}

\subsection{Simulation Setup}
The short-lived NS remnant we evolve formed in the merger of an equal-mass binary with individual NS masses of 1.35$M_{\odot}$ at infinity, originally simulated in GRHD by \cite{Radice2018} using the \texttt{WhiskyTHC} code. We map the NS remnant as initial data to our simulation at 17 ms post-merger, adding a poloidal magnetic field of strength $B_0$=10$^{15}$ G. Our simulation employs ideal GRMHD using the \texttt{Einstein Toolkit} and includes the $K_0 = 220$ MeV variant of the equation of state of \cite{Lattimer1991}. Neutrino-matter interactions are tracked using a recently developed M1 neutrino transport scheme \citep{Radice2022}. We initialize neutrino quantities by assuming equilibrium conditions. In \cite{Moesta2020}, the same NS remnant was evolved using a similar setup, but using a leakage scheme to treat neutrinos. 

\subsection{Tracer Particles and Nuclear Reaction Network}
\label{sec:methods_nucysn} 

We use tracer particles to extract the thermodynamic conditions of the ejected material. The tracers are uniformly spaced to represent regions of constant volume. At the start of the simulation, each tracer particle is assigned a mass that accounts for the density at its location and the volume the particle covers. We place 96000 tracer particles to ensure that a sufficient number of tracer particles are present in the outflow. The tracers are advected with the fluid and collected at a surface defined by a chosen radius, $r=$150 M$_{\odot}$ $=222$ km here, and tracer quantities are frozen once the tracer particle crosses this surface.

We compute the ejecta composition by post-processing the tracers with the \texttt{SkyNet} nuclear reaction network \citep{Lippuner2017}. We include 7852 isotopes up to $^{337}$Cn. The reaction rates, nuclear masses and partition functions are the same as those used in \cite{Curtis2023}. The network is started in NSE when the particle temperature drops below 20~GK. The network calculates the source terms due to individual nuclear reactions and neutrino interactions, and evolves the temperature. The dynamical simulation, and hence the tracer trajectory, ends around 12 milliseconds after the mapping due to the collapse of the remnant. The network continues the nucleosythesis calculation up to a desired end time by smoothly extrapolating the tracer data beyond the end of the trajectory, assuming homologous expansion. Our calculations are carried out to 10$^9\,$s, long enough to generate a stable abundance pattern as a function of mass number.

\subsection{Radiation Transport}

We compute kilonova light curves using \texttt{SNEC} \citep{Morozova2015, Wu2022}, a spherically-symmetric Lagrangian radiation hydrodynamics code capable of simulating the hydrodynamical evolution of merger ejecta and the corresponding bolometric and broadband light curves. As input, \texttt{SNEC} requires the radius, temperature, density, velocity, initial $Y_e$, initial entropy and expansion timescale of the outflow as a function of mass coordinate. The $Y_e$, entropy and expansion timescale are used to compute the heating rates and opacities. The outflow properties are recorded by measuring the flux of the relevant quantities through a spherical surface at radius $r=$ 100$M_{\odot}$ $=148$ km. We use the same approach as employed in \cite{Curtis2023}.

\section{Results}
\label{sec:results}

\subsection{Outflow Properties}
\label{sec:hmns:outflows}

\begin{figure*}

\includegraphics[width=\textwidth]{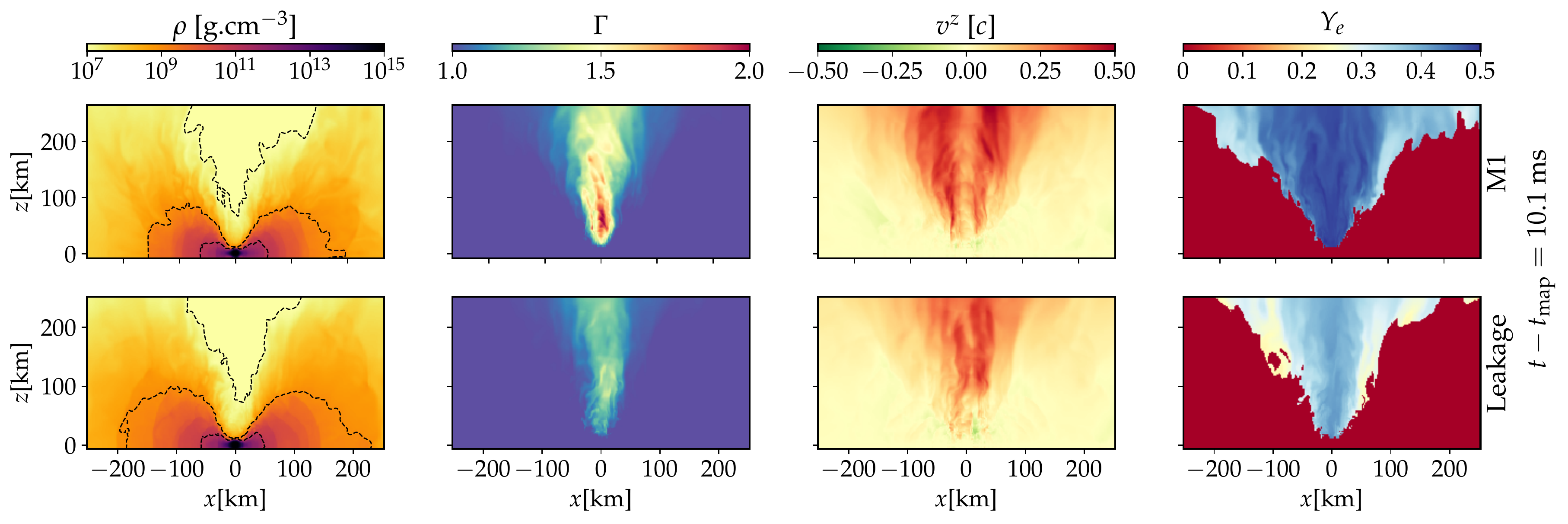}

\caption{ Meridional slices, \emph{i.e.} $xz$-plane with $z$ being the vertical axis, of the density $\rho$, the Lorentz factor at infinity $\Gamma$, the velocity component aligned with the rotation axis $v^z$ and the electron fraction $Y_e$. We only plot the $Y_e$ for the unbound material. We show snapshots at $t-t_{\mathrm{map}}=10.1$ms, where $t_{\mathrm{map}}=17\ \mathrm{ms}$ after coalescence of the neutron star binary. The top row shows the evolution of the remnant using an M1 scheme while the bottom row shows a similar evolution using a leakage scheme. The dashed are isodensity contours of $\rho= 10^7,\ 10^8,\ 10^9,\ 10^{11}\ \mathrm{g~ cm}^{-3}$.
}\label{fig:hydro}
\end{figure*}

We present 2D snapshots of the simulation at t=10.1ms in Figure \ref{fig:hydro}, along with the corresponding snapshots from the simulation presented in \cite{Moesta2020}. The simulations differ in their treatment of neutrino physics -- while the simulation in this paper employs an M1 transport scheme, the previous work used a leakage scheme. The simulation presented here leads to an earlier remnant neutron star collapse compared to our previous work (12ms vs 22ms) but the overall outflow geometry and properties are very similar. The ejecta are concentrated in the polar region, as can be seen in the panels depicting the $Y_e$ of the unbound material in Figure \ref{fig:hydro}. Roughly $\sim 3 \times 10^{-3} M_{\odot}$ of material becomes unbound during the course of our simulation. The mass ejection rate for the M1 simulation when the outflow has reached a steady-state is $\simeq 0.08\, M_{\odot}\, s^{-1}$, roughly consistent with our previous results.   

\begin{figure*}
\begin{center}
        \begin{tabular}{ccc}
        \includegraphics[width=0.33\textwidth]{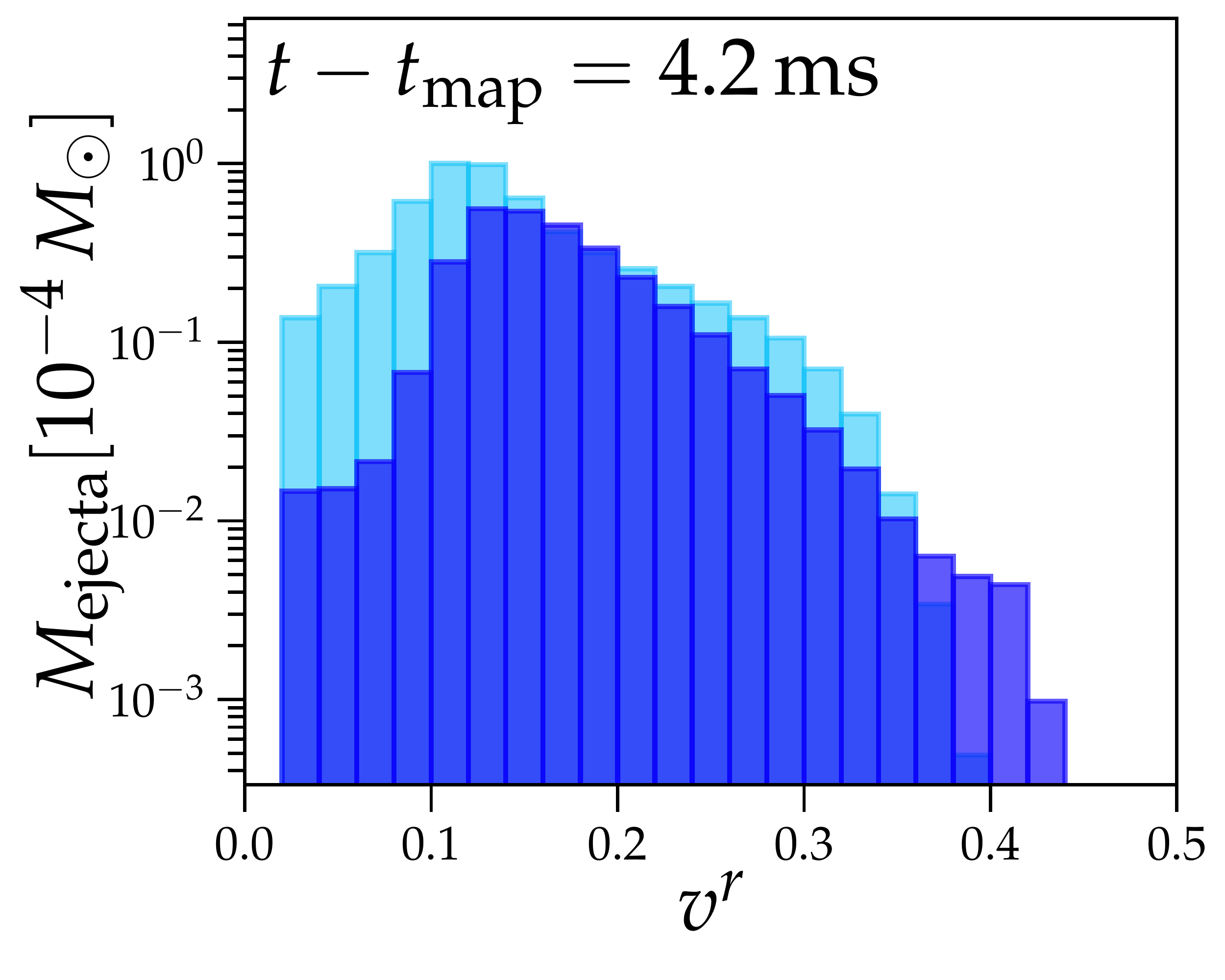}
        \includegraphics[width=0.33\textwidth]{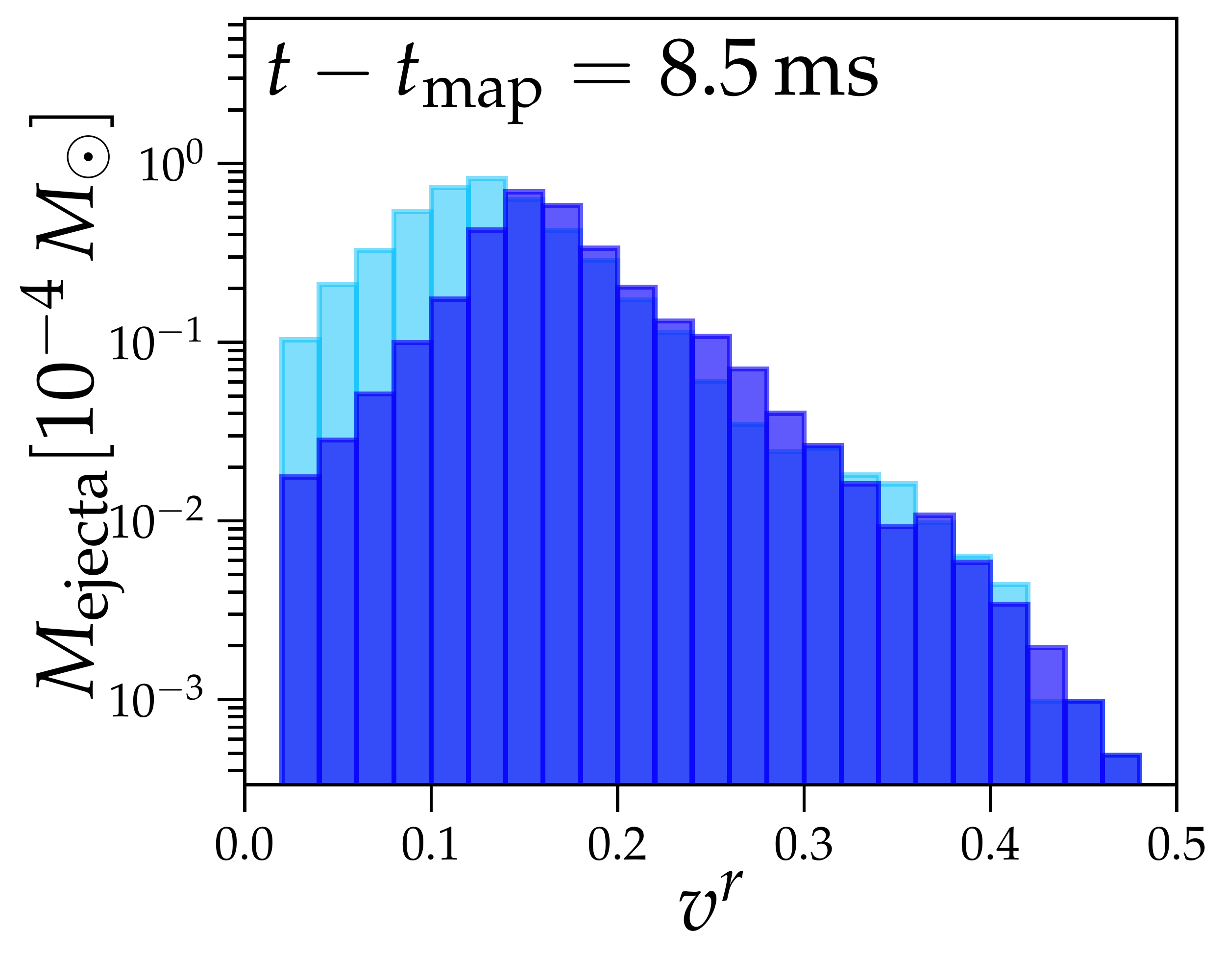}
        \includegraphics[width=0.33\textwidth]{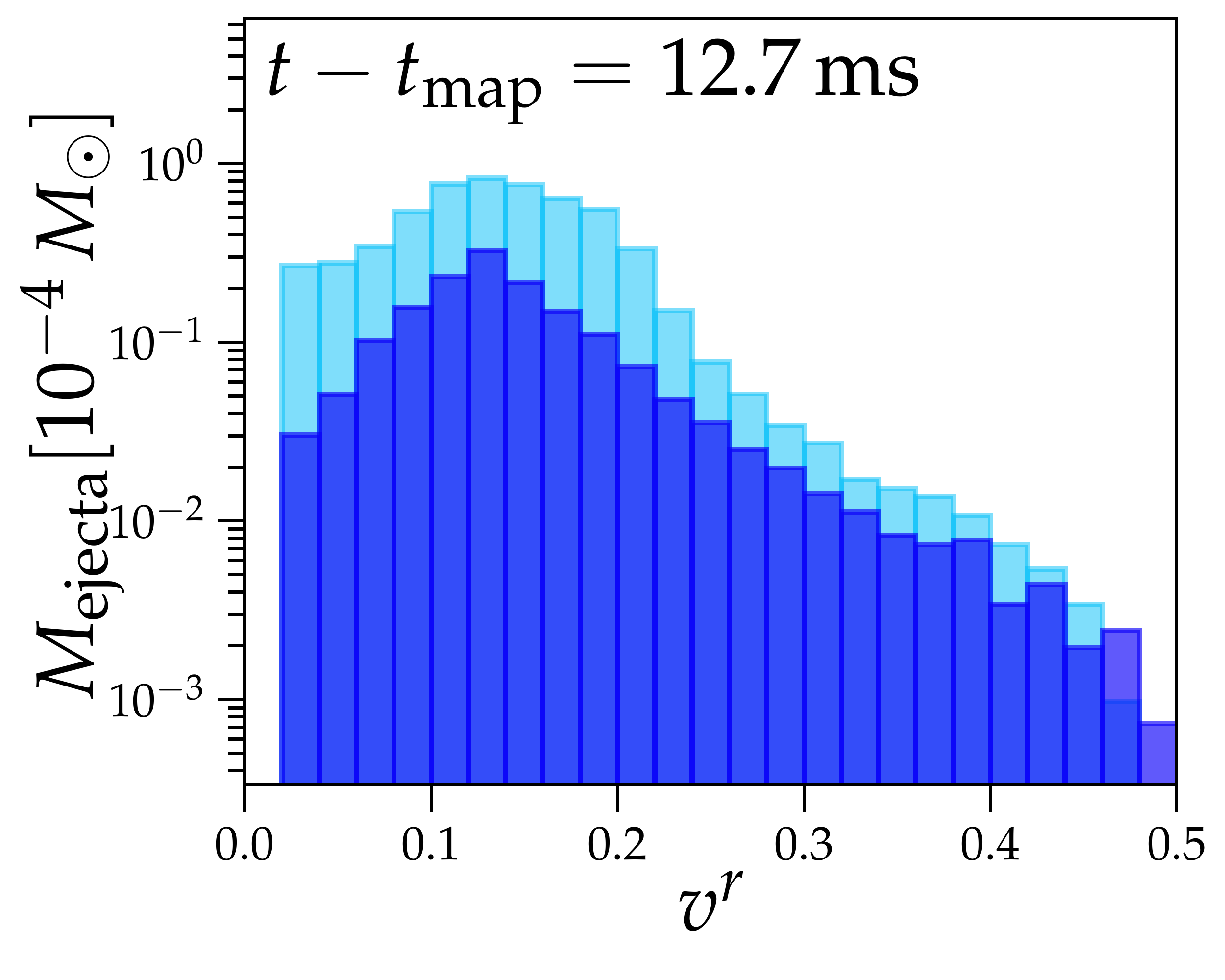}\\

        \includegraphics[width=0.33\textwidth]{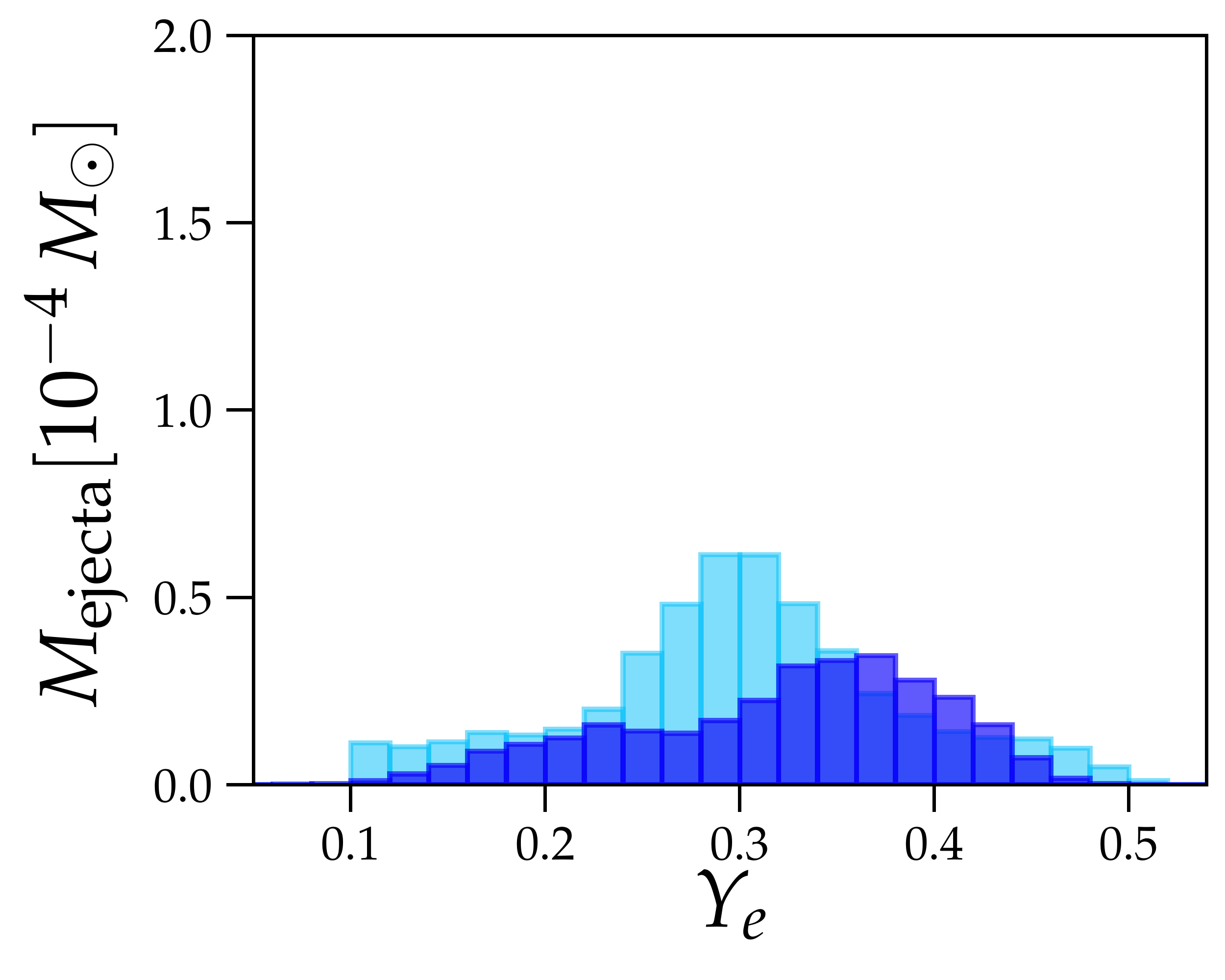}
        \includegraphics[width=0.33\textwidth]{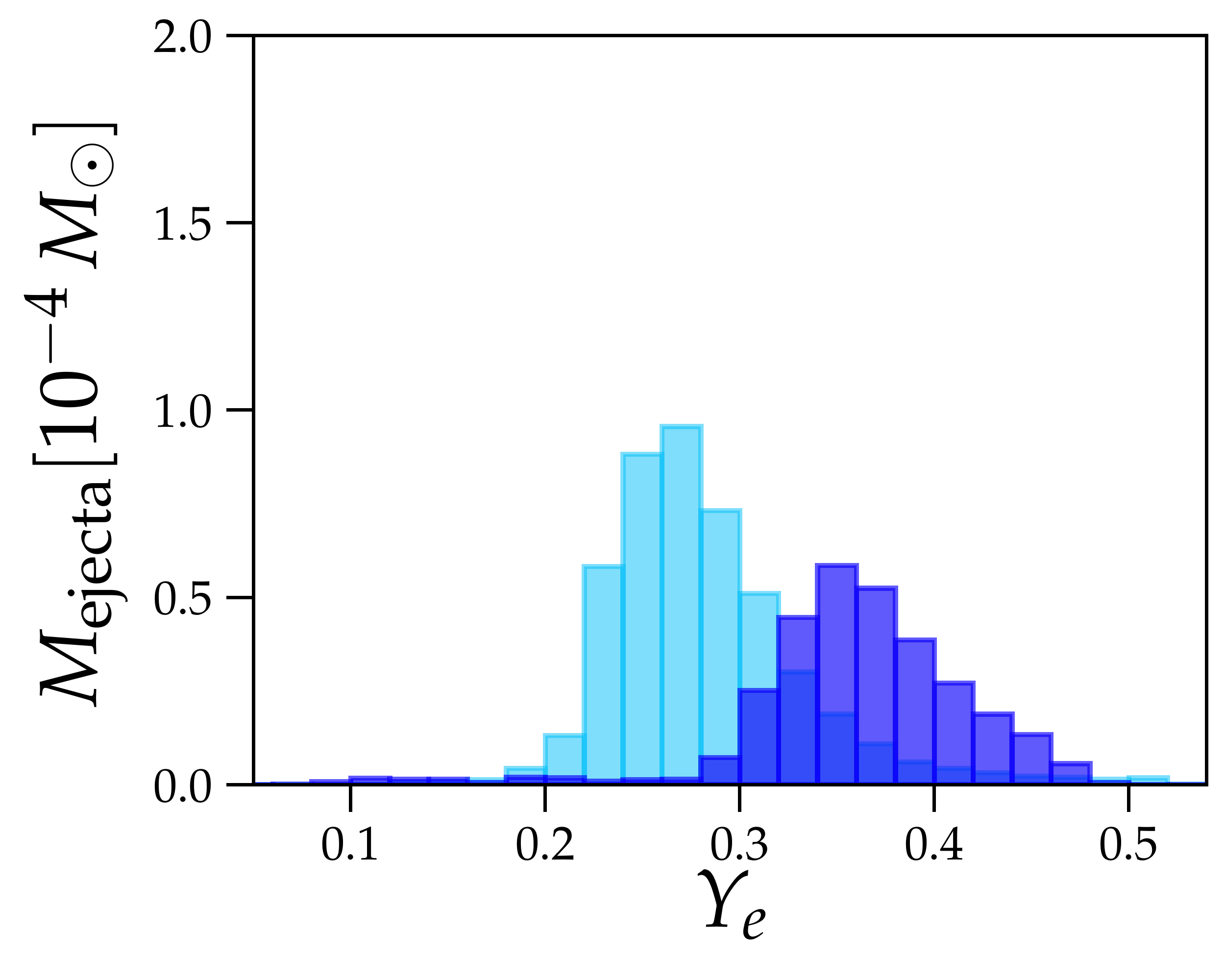}
        \includegraphics[width=0.33\textwidth]{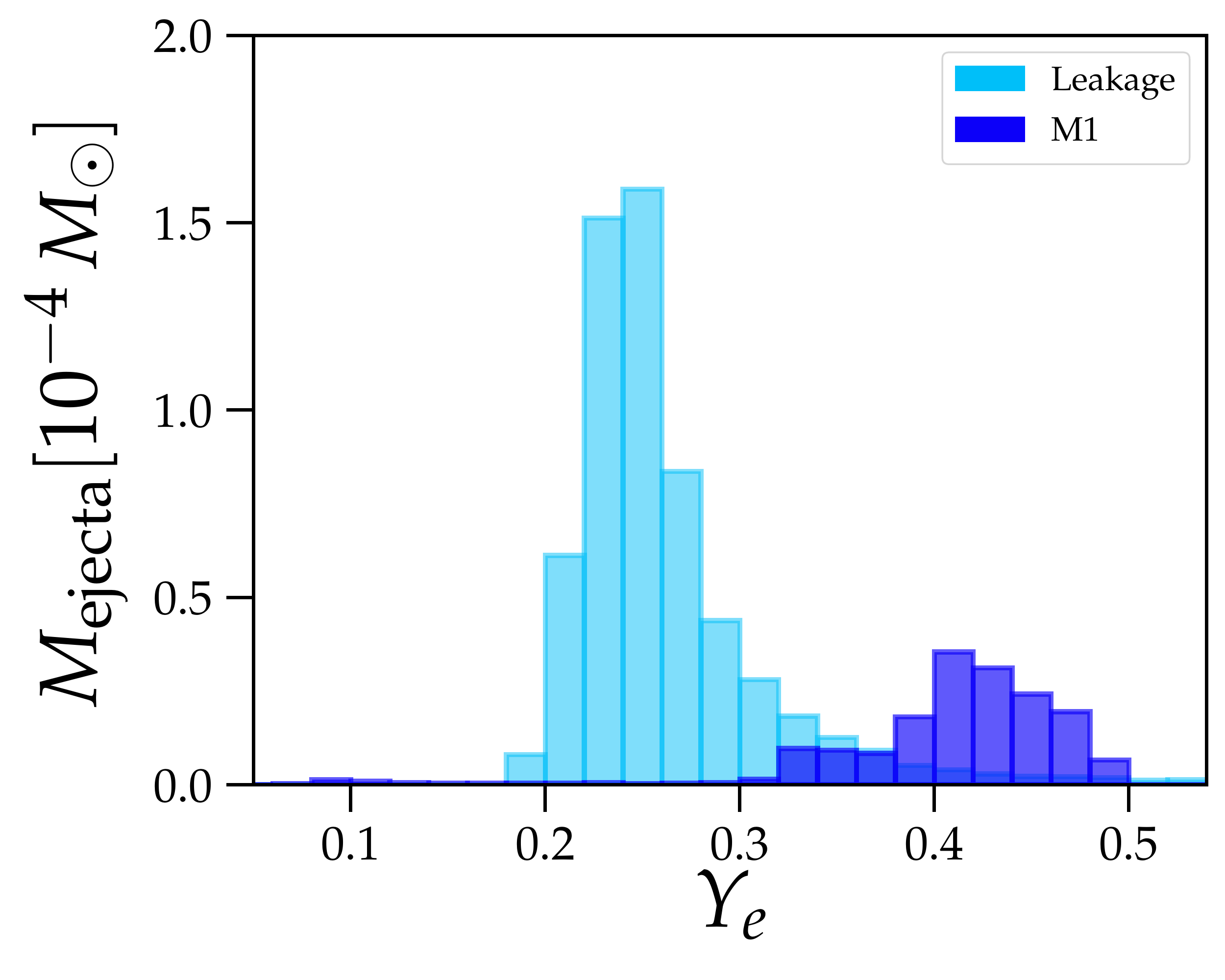}

\end{tabular}
	\caption{Histograms of the radial velocity $v^r$ of the unbound material (top panels), where $r$ is the radius in spherical coordinates, and the $Y_e$ of the unbound material (bottom panels), shown at different times during the dynamical simulation (4.2, 8.5 and 12.7ms), for the leakage and M1 simulations. Each column corresponds to a different time in the evolution.  
 }.
        \label{fig:vr_hist}
\end{center}
\end{figure*}

In Figure \ref{fig:vr_hist}, we present histograms of the radial velocity of the unbound material and its electron fraction at various times during the two simulations. The unbound material is determined via the Bernoulli criterion. The ejecta show a broad, and overall very similar velocity distribution. A significant amount of material has velocities between $0.2c < v^r < 0.3c$ at all times, consistent with the range of velocities estimated for the `blue' ejecta component associated with AT2017gfo. As the system evolves, a small fraction of the ejecta attains velocities in the range $0.4c < v^r < 0.5c$. 

Examining the evolution of the $Y_e$ distributions, we note the effect of neutrino-matter interactions in the M1 simulations, which drive the ejecta $Y_e$ towards higher values over time, as discussed in some detail in \cite{Radice2022} and \cite{Zappa2023}. In the leakage simulation, most of the ejected material has $Y_e$ values between 0.2 -- 0.3, with a peak around $Y_e \sim 0.25$, but the M1 simulation predicts meaningfully higher ejecta $Y_e$. The leakage scheme used in \cite{Moesta2020} included an approximate energy deposition rate due to neutrino absorption in optically thin conditions, but did not directly include the effect of neutrino absorption on the composition. The inclusion of this effect in the M1 simulation results in the increase in $Y_e$ observed. In the M1 simulation, barely any of the material ejected at later times has $Y_e \lesssim 0.3$. 

While the two snapshots in time do not exactly correspond to the same stage in the simulation (due to the different collapse times of the remnant) the difference in outflow composition persists across the entire simulated time. The systematically higher $Y_e$ values found using the M1 scheme as compared to leakage agree with the results of \citep{Radice2022}. This range of $Y_e$ values in the outflow is expected to inhibit the synthesis of a significant mass-fraction of lanthanides. Given the typical entropies and dynamical timescales found in these simulations, the lanthanide turn-off point is expected to be $Y_e \sim 0.24$ \citep{Lippuner2015}.

\subsection{r-Process Nucleosynthesis}
\label{sec:hmns:r-process}

\begin{figure}
\begin{center}
        \begin{tabular}{c}
        \includegraphics[width=0.45\textwidth]{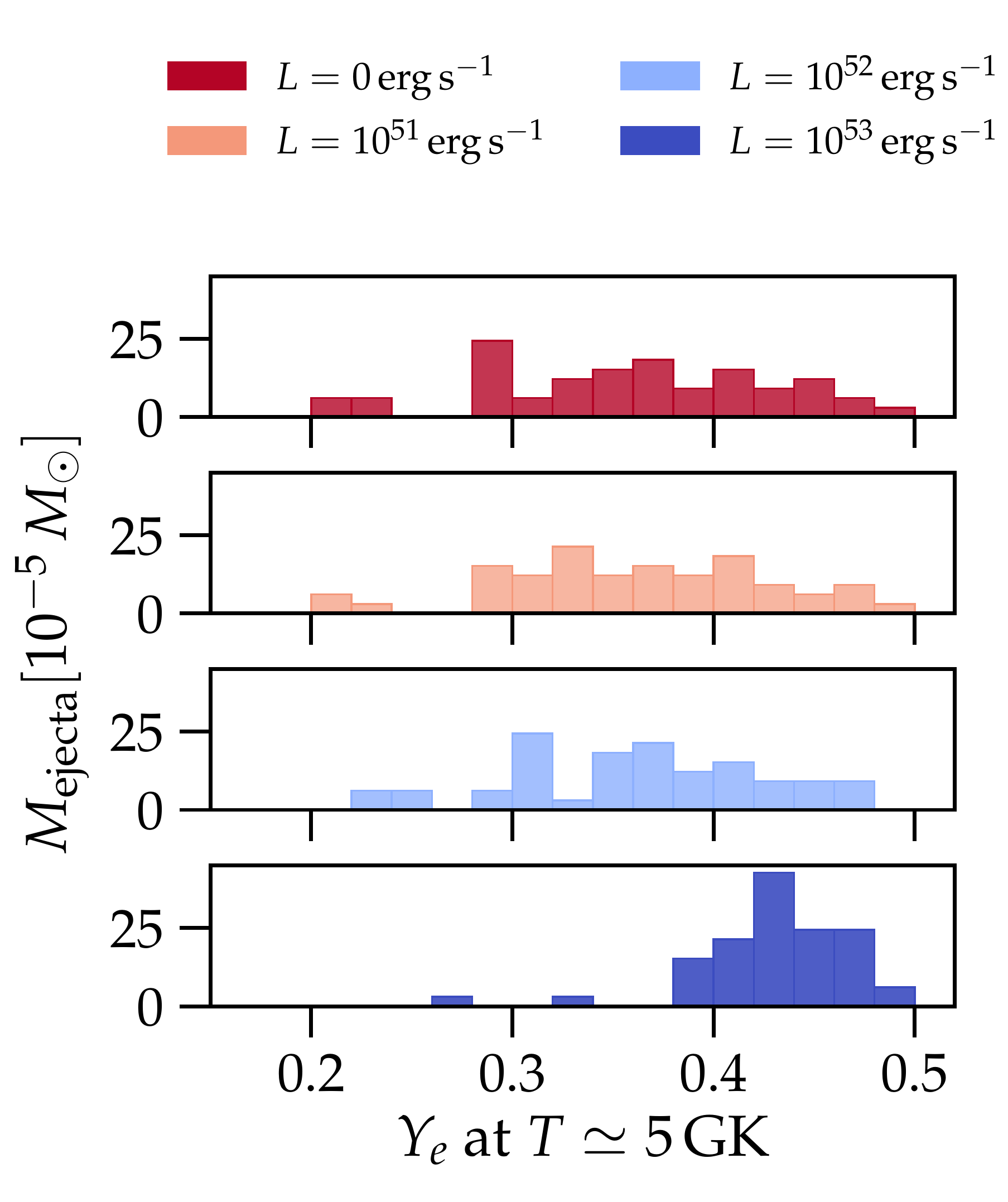}\\
        \includegraphics[width=0.45\textwidth]{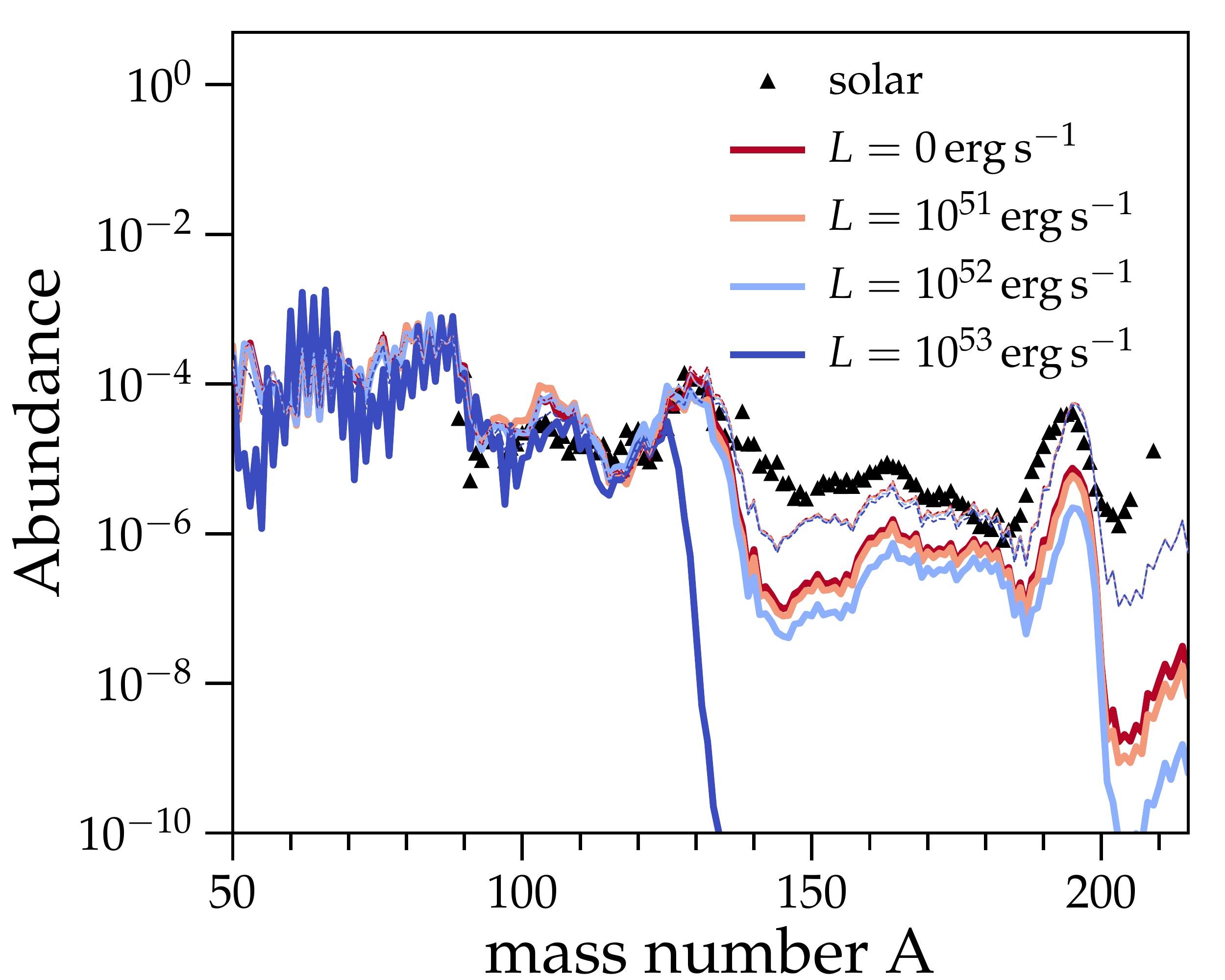}\\
\end{tabular}
	\caption{Mass-weighted abundance as a function of mass number for the NS remnant outflow. The different colors correspond to different constant neutrino luminosities employed during post-processing. The solar abundance pattern has been scaled to match the second $r$-process peak at $A \sim 130$ for the zero luminosity setting. The dashed lines show the total abundance pattern produced in combination with the dynamical ejecta from \cite{Radice2018}}.
        \label{fig:abun}
\end{center}
\end{figure}

The mean outflow entropy is roughly 20 $k_{\rm B}/{\rm baryon}$ with typical expansion timescales of the order $\sim 10-20$ ms. In the top panel of Figure \ref{fig:abun},  we show the $Y_e$ distribution for all ejected tracers when the temperature of the particles is last above 5~GK (the relevant temperature for $r$-process nucleosynthesis), as computed within \texttt{SkyNet}. This is distinct from the ejecta $Y_e$ shown in Figure \ref{fig:vr_hist} since the $Y_e$ is evolved within the network based on the neutrino luminosity employed during post-processing.

We plot the $Y_e$ distribution for four calculations, each assuming a different constant value of the neutrino luminosity, ranging from $0 - 10^{53}$ erg s$^{-1}$. This range of constant neutrino luminosities is used to bracket the possible uncertainties in composition arising from the approximate neutrino transport treatment. For the constant luminosity calculations, we assume $L_{\nu_e} = L_{\nu_{\bar{e}}}$ and constant mean neutrino energies $\langle \epsilon_{\nu_e}\rangle = 10\, \mathrm{MeV}$ and $\langle \epsilon_{\bar{\nu_e}} \rangle = 14\, \mathrm{MeV}$. The luminosities observed in the simulation are a few $10^{52}\, \mathrm{erg}\, \mathrm{s}^{-1}$, in agreement with what \citep{Cusinato2022} found. We will present the neutrino properties in the simulation in a future study.

In general, higher constant neutrino luminosities shift the $Y_e$ distribution towards higher values, approaching the equilibrium $Y_e$. The exact value of $Y_e$, and whether weak equilibrium is actually attained, is decided by the competition between the weak interaction timescale and the dynamical timescale. Here, most of the ejected material has $Y_e \gtrsim 0.3$ for all calculations, and for the extreme case where $L_\nu = 10^{53}$ erg s$^{-1}$, the ejecta $Y_e$ peaks around $\sim$0.44. 

In the bottom panel of Figure \ref{fig:abun}, we show the corresponding abundance patterns. The $r$-process abundances produced do not match the solar pattern for any of the scenarios, and the abundances of heavy nuclei are increasingly suppressed as we increase the constant neutrino luminosity employed during post-processing. In combination with the dynamical ejecta, the solar pattern can be reproduced for the simulation presented here, however, for longer-lived remnants, we expect the contribution of the remnant outflows to dominate the total abundance pattern. The lanthanide fractions for the four constant luminosity runs, in order of increasing $L_\nu$, are 3.1 $\times 10^{-3}$, 2.6 $\times 10^{-3}$, 1.4 $\times 10^{-3}$ and 2.13 $\times 10^{-10}$. 

Typically, for $X_{\textrm{La}} \gtrsim 3\times 10^{-3}$, the kilonova peaks in the near-infrared J band around 1 day \citep{Even2020}. However, since the $Y_e$ of our outflows (and hence the ejecta composition) varies significantly with angle off of the midplane, the character of the kilonova observed will depend heavily on the viewing angle, in agreement with various previous studies based on numerical relativity simulations, radiative transfer simulations, and Bayesian analyses of observational data, e.g. \citep{Perego2017, Kawaguchi2018,Breschi2021MNRAS}.

\subsection{Blue Kilonova}
\label{sec:hmns:kn}

\begin{figure*}
\begin{center}
    \begin{tabular}{cc}
        \includegraphics[width=0.48\textwidth]{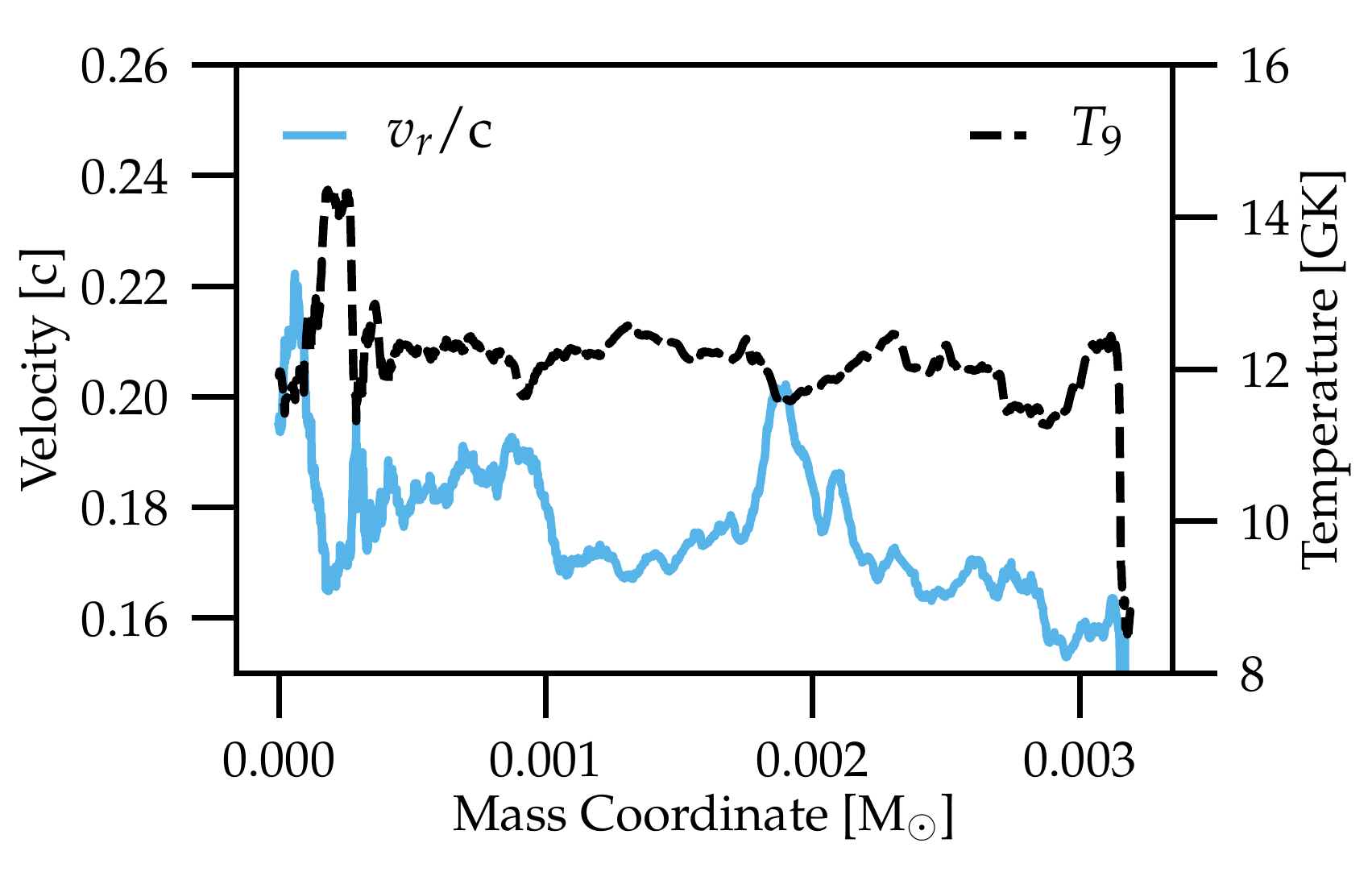}     \includegraphics[width=0.48\textwidth]{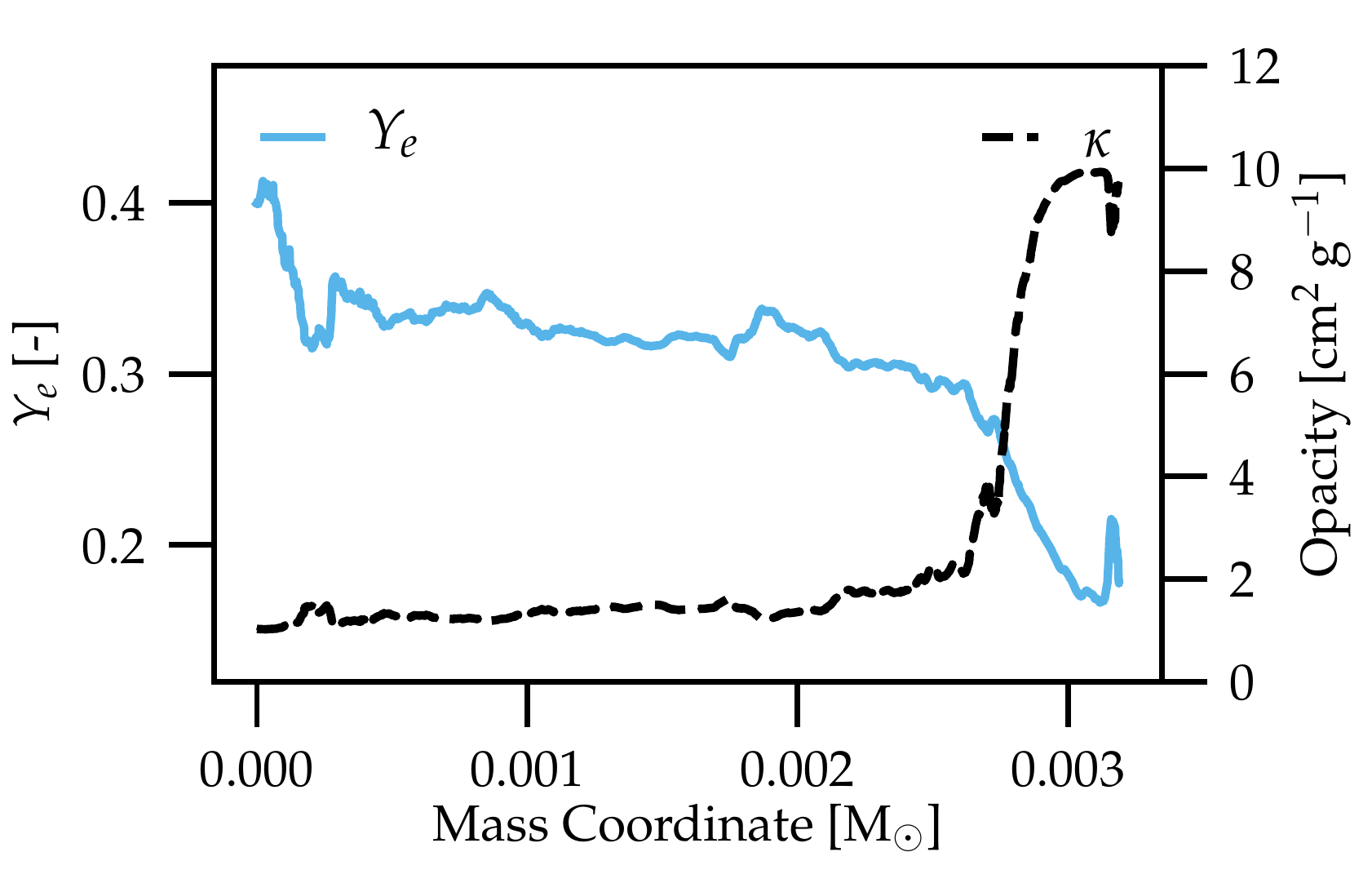}      
         \\
        \includegraphics[width=0.48\textwidth]{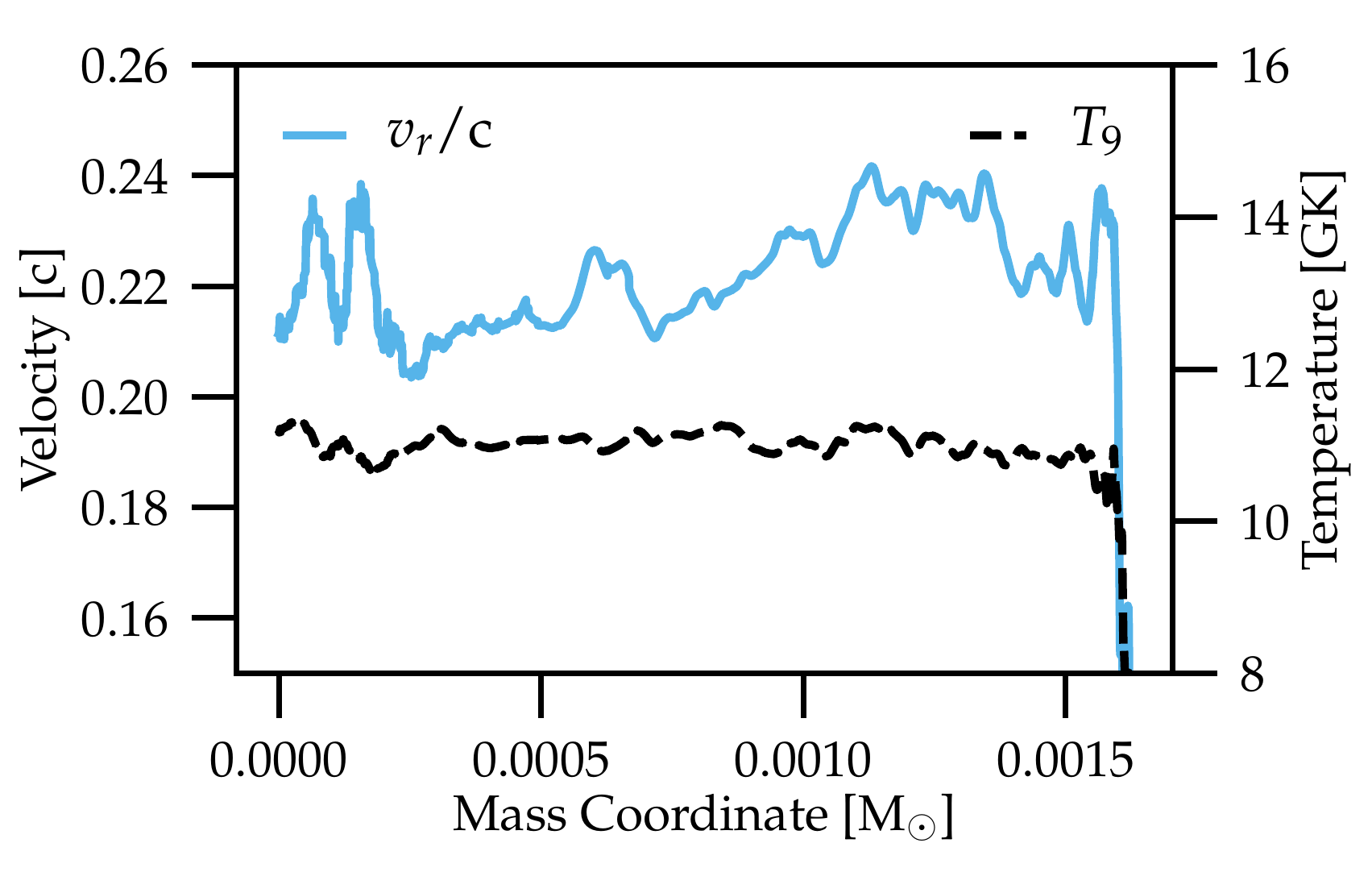}    \includegraphics[width=0.48\textwidth]{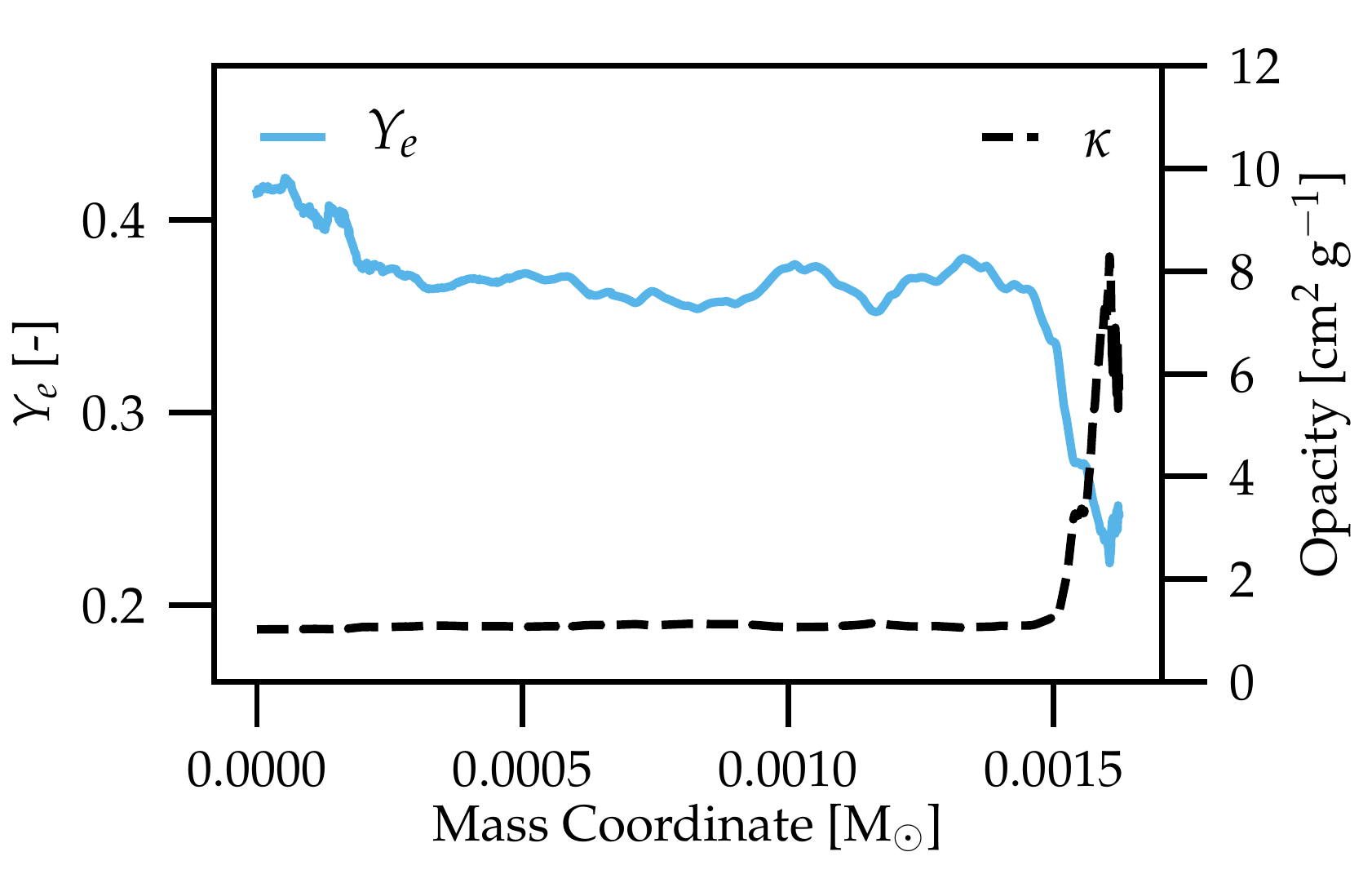}
       
    \end{tabular}  
	\caption{Input profiles for SNEC showing averaged quantities for the NS remnant outflow as a function of mass coordinate. The top two panels show the radial velocity (solid blue line) and temperature (dashed black line) of the total ejecta on the left, and the corresponding $Y_e$ (solid blue line) and opacity (dashed black line) on the right. The bottom two panels show the same quantities plotted for the polar ejecta.} 
        \label{fig:snec_input}
\end{center}
\end{figure*}

We calculate the broadband magnitudes of the kilonova using the \texttt{SNEC} code, which models outflows as spherically-symmetric. However, the actual ejecta morphology is complex. The ejecta $Y_e$, which directly sets the opacity in our radiation transport treatment, shows a significant dependence on latitude, with relatively low values closer to the equator and values approaching 0.5 closer to the poles, as seen in Figure \ref{fig:hydro}. This introduces a viewing-angle dependence of the kilonova which cannot be adequately captured by the averaged outflow profile. We therefore compute the averaged profile of the ejecta in the polar region in addition to the averaged profile of the total ejecta.
 
Figure \ref{fig:snec_input} presents the velocity, temperature, $Y_e$ and opacity profiles of the NS remnant outflow used as input for \texttt{SNEC} calculations. The two rows correspond to the averaged profiles of the total ejecta and the polar ejecta (contained within 42$^{\circ}$ as measured from the poles). The $Y_e$ of the ejecta increases systematically as we move inward in mass coordinate, reflecting the increase in the $Y_e$ of the ejected material over time. Examining the total ejecta profile, we find temperatures around 12~GK and velocities between 0.16--0.2$c$. Most of the ejecta have $Y_e \gtrsim 0.3$, but a small amount of material present in the outermost layers of the total ejecta profile has $Y_e \lesssim$ 0.2. This material was ejected at early times at lower latitudes. Our opacity prescription assigns opacity values $\sim$10 cm$^2$ g$^{-1}$ to the outermost, low-$Y_e$ layer of the ejecta, dropping as we move inward in mass coordinate to $\sim 1-2$ cm$^2$ g$^{-1}$ for the bulk of the total ejecta with $Y_e \gtrsim$ 0.3. The outermost high opacity ejecta may serve as a lanthanide curtain \citep{Kasen2015, Wollaeger2018, Nativi2021}, absorbing blue light and shifting the kilonova peak to redder bands. Examining the polar ejecta only, we find temperatures around 11~GK and relatively higher velocities between 0.20--0.24$c$. The corresponding profile shows distinctly higher $Y_e$ than the profile for the total ejecta, with typical $Y_e\gtrsim 0.35$. Most importantly, the lanthanide curtain nearly absent, given most of the low $Y_e$ material which absorbs blue light lies close to the equator.

\begin{figure*}
\begin{center}
 \includegraphics[width=1.0\textwidth]{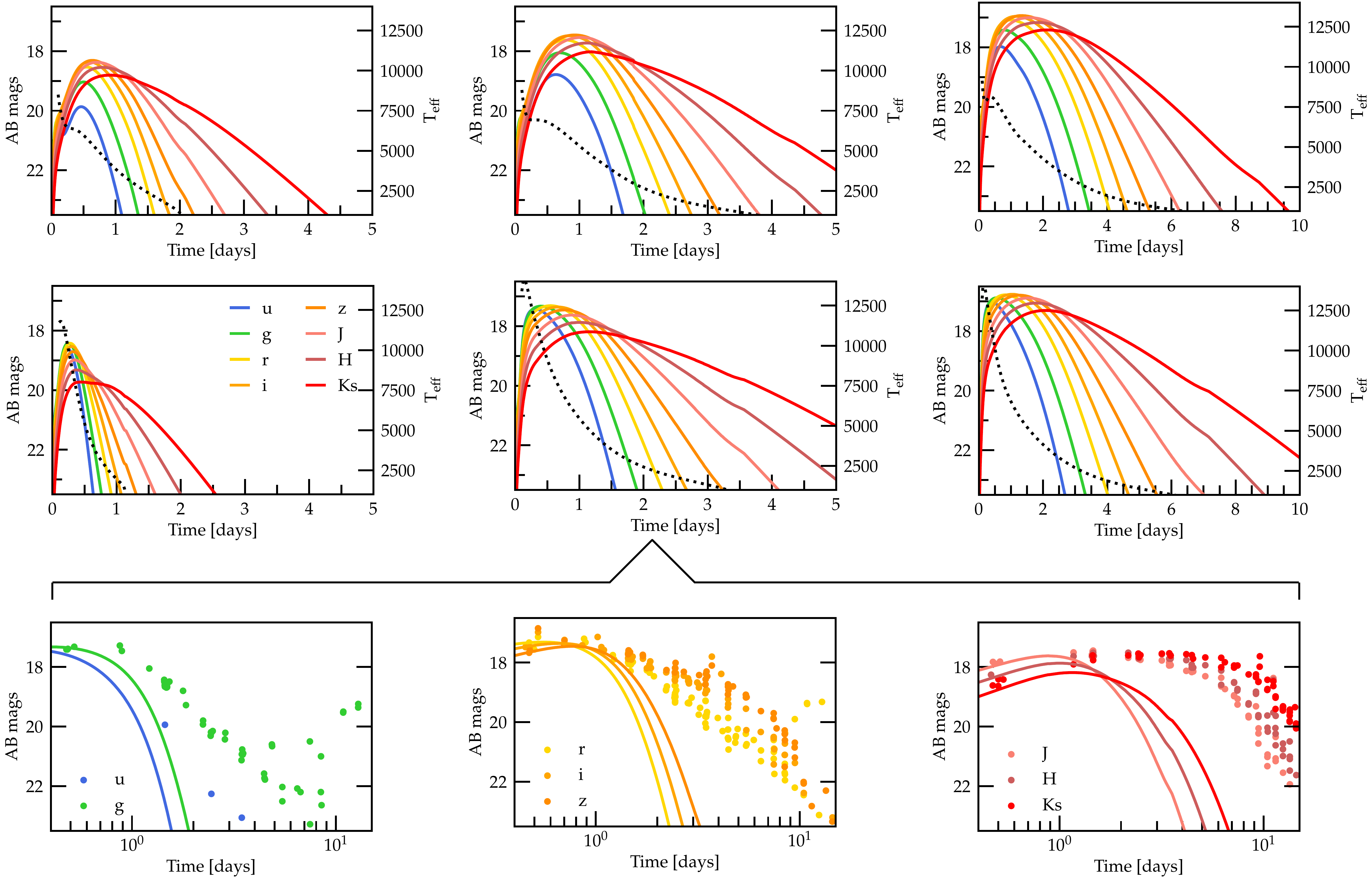}
 \caption{AB magnitudes in $ugrizJHK_{s}$ bands. Top two rows: band light curves for the total ejecta and the polar ejecta. From left to right, the band light curves correspond to the ejecta produced during the course of the simulation, and those produced assuming a total remnant neutron star lifetime of $\sim$240ms and $\sim$1s, respectively. Bottom row: comparison with broadband data for AT2017gfo, for the case where we obtain $\sim$0.02$M_{\odot}$ of `blue' ejecta consistent with the blue kilonova component. The aggregated broadband data plotted here were collected by multiple teams and compiled in \cite{Villar2017}.}
        \label{fig:snec_kn}
\end{center}
\end{figure*}

In Figure \ref{fig:snec_kn}, we present the AB magnitudes in optical ($ugriz$) and near-infrared (JHK$_s$) filters, computed under the assumption of blackbody emission at the photosphere and for the layers above the photosphere. The distance between the observer and the kilonova is taken to be 40 Mpc, same as the approximate distance to AT2017gfo. The black dotted line gives the evolution of the effective temperature computed at the photosphere. We present light curves for the total ejecta and the polar ejecta, for the present simulation as well as under the assumption of continued mass ejection by a longer-lived NS remnant.

For the total ejecta profile, the brightest emission is seen in the $i$ and $z$ bands. Due to the presence of high opacity material in the outer layers, by the time these ejecta expand enough to radiate efficiently, they have also cooled down substantially. The majority of the observed emission therefore happens at longer wavelengths. For the polar ejecta, where the outermost layer of lanthanides is nearly non-existent, the light curve peaks sooner and the brightest emission seen in the $g$ and $r$ i.e. `blue' optical bands. If we consider all ejecta within a polar angle of 54$^{\circ}$, then the kilonova peaks in the $r$ and $i$ bands instead. 

The total ejecta mass in our simulations is an order of magnitude lower than that estimated for the blue component of AT2017gfo owing to the short lifetime of the remnant. For longer-lived remnants ($O \sim 100$ ms), we expect the mass ejection to continue at the quasi-steady-state rate until the NS remnant collapses. We explore this possibility using extrapolated ejecta profiles assuming the appropriate mass ejection rates and mean values for the relevant physical quantities. The resulting kilonovae are also shown in Figure \ref{fig:snec_kn}. We find a range of outcomes, both respect to kilonova peak magnitude, timescale, and color. For the case where we obtain $\sim 2 \times 10^{-2} M_{\odot}$ of ejecta, which is the total mass inferred for the blue component of AT2017gfo, the blue kilonova produced is broadly consistent with the early blue emission observed for AT2017gfo, as is shown in the bottom row of Figure \ref{fig:snec_kn}. The late time behavior arises from other ejecta components. 

\section{Summary and Discussion}
\label{sec:summary}
We present a 3D GRMHD simulation of a short-lived NS remnant formed in the aftermath of a BNS merger. The simulation employs a microphysical finite-temperature EOS and an M1 scheme for neutrino transport. A magnetized wind is launched from the short-lived NS remnant and ejects neutron-rich material with a rate $0.8 \times 10^{-1}\ M_{\odot} {\rm s^{-1}}$. We compute nucleosynthesis yields in these ejecta using tracer particles post-processed with \texttt{SkyNet}. We also record outflow properties and map them to the \text{SNEC} code in order to predict the resulting kilonova.

The ejecta $Y_e$ distribution is broad, similar to our previous work employing a leakage scheme, as well as other studies in the literature. However, the distribution peaks at relatively high values $Y_e \gtrsim 0.3$ at early times, and the peak shifts to $Y_e \gtrsim 0.4$ at later times\citep{Radice2022, Zappa2023}. Such high $Y_e$ values, in combination with the high velocity and mass ejection rate, make magnetically-driven NS remnant ejecta (see \cite{Moesta2020} for the importance of the inclusion of MHD for these outflows) a distinct and potentially dominant component of merger ejecta, and a promising source of blue kilonovae. The bulk of the dynamical ejecta have much lower $Y_e$, albeit a comparable total mass, while the disk winds have much lower velocities than the NS remnant ejecta. 

The production of $r$-process elements is suppressed in these ejecta relative to the solar pattern. Depending on the neutrino luminosity employed during post-processing, the ejecta lanthanide fractions can vary between 3.1 $\times 10^{-3}$ to a negligible amount. However, combining the ejecta produced during the course of our simulation and the dynamical ejecta will result in an abundance pattern that is close to solar.

The kilonova color depends on the viewing angle due to the changing ejecta composition as a function of latitude \citep{Perego2017, Kawaguchi2018,Breschi2021MNRAS}. The lower-$Y_e$ material present at lower latitudes contains a larger mass-fraction of lanthanides, while production of these isotopes is suppressed in the high-$Y_e$ ejecta closer to the poles. Taking the total ejecta into account in a spherically-symmetric radiation hydrodynamics simulation, we find peak magnitudes in the $i$ and $z$ bands. However, for the polar ejecta, we find a transient that peaks in the bluer $g$ and $r$ bands. 

For longer-lived NS remnants, significantly more mass can be ejected by the wind. Here, we find that a remnant with a lifetime of $\sim$ 240ms can produce a blue kilonova compatible with AT2017gfo. How long the NS must survive in order to produce sufficient ejecta is estimated based on the mass ejection rate, and is therefore sensitive to the binary properties, the NS equation of state, and the physics implemented in the simulation. This is the first demonstration of the production of blue kilonovae compatible with AT2017gfo from NS remnant outflows based on high physical-fidelity GRMHD simulations, and including M1 neutrino transport as well as finite-temperature equation of state effects.

The dynamical merger ejecta synthesize substantial amounts of lanthanides \citep{Radice2016MNRAS, Kullmann2022, Radice2022, Just2023arXiv}. Thus, in combination with the NS remnant ejecta, a kilonova with both blue and red components can be produced. While the dynamical ejecta dominate the early emission in the equatorial region, they are not expected to obscure the kilonova produced by the NS remnant outflows. The remnant neutron-star outflows are also much faster than post-merger accretion disk winds, and thus the production of an early blue kilonova should not be affected by disk-wind component launched after the NS remnant collapses to a black hole.  

In this work, we have produced kilonova light curves for NS remnant ejecta under the assumption of spherical symmetry. In a follow up work, we will model a longer-lived NS remnant with properties similar to GW170817 and perform multi-dimensional kilonova calculations, tracking ejecta evolution from the moment of launch to the kilonova phase. Our work complements ongoing efforts to model populations of kilonovae and determine observing strategies for upcoming surveys such as LSST \citep{Setzer2023, Ekanger2023arXiv}. Such efforts need to take the NS remnant outflows into consideration in order to capture the full spectrum of kilonova transients.

\section{Acknowledgements}
We thank Dan Kasen for useful discussions and Ashley Villar for providing the compiled data for AT2017gfo. 
PM and PB acknowledge funding through NWO under grant number OCENW.XL21.XL21.038.
DR~acknowledges funding from the U.S. Department of Energy, Office of Science, Division of Nuclear Physics under Award Number(s) DE-SC0021177 and from the National Science Foundation under Grants No. PHY-2011725, PHY-2020275, PHY-2116686, and AST-2108467.
SB acknowledges support by the EU Horizon under ERC Consolidator Grant, no. InspiReM-101043372 and from the Deutsche Forschungsgemeinschaft (DFG) project MEMI number BE 6301/2-1.
Research at Perimeter Institute is supported in part by the Government of Canada through the Department of Innovation, Science and Economic Development and by the Province of Ontario through the Ministry of Colleges and Universities.
RH acknowledges funding through NSF grants OAC-2004879,  OAC-2103680, and OAC-1238993.
The simulations have been carried out on Snellius at SURF and SuperMUC-NG at GCS@LRZ, Germany. We acknowledge PRACE for awarding us access to SuperMUC-NG at GCS@LRZ, Germany.

\bibliography{main.bib}{}
\bibliographystyle{aasjournal}


\end{document}